\documentclass[prb,a4paper,10pt,twocolumn,showpacs,floatfix,superscriptaddress,amsmath,amsfonts,amssymb,preprintnumbers,longbibliography]{revtex4-1}
\usepackage[utf8]{inputenc} 
\usepackage{float}
\usepackage{dcolumn,graphicx,color,booktabs,microtype,afterpage}
\graphicspath{{./}{figure/}}
\usepackage[charter,greekuppercase=italicized]{mathdesign}
\usepackage{sidecap}
\renewcommand{\tablename}{Table}
\makeatletter\renewcommand{\fnum@figure}[1]{\figurename~\thefigure.~(Color online)~}\makeatother
\makeatletter\renewcommand{\fnum@table}[1]{\tablename~\thetable.}\makeatother

\newcount\hh \newcount\mm
\hh=\time \divide\hh by 60
\mm=\hh \multiply\mm by 60 \mm=-\mm
\advance\mm by \time
\def\now{\number\hh:\ifnum\mm<10{}0\fi\number\mm}

\usepackage[colorlinks,plainpages=false,linkcolor=blue,urlcolor=blue,citecolor=blue,pdfpagemode=UseNone,pdfstartview=FitBH]{hyperref}

\usepackage{nicefrac}
\newcommand{\CUPZ}{Cu\-(pz)$_2$\-(ClO$_4$)$_2$}

\begin{document}

\makeatletter\renewcommand{\ps@plain}{%
\def\@evenhead{\hfill\itshape\rightmark}%
\def\@oddhead{\itshape\leftmark\hfill}%
\renewcommand{\@evenfoot}{\hfill\small{--~\thepage~--}\hfill}%
\renewcommand{\@oddfoot}{\hfill\small{--~\thepage~--}\hfill}%
}\makeatother\pagestyle{plain}


\title{~\vspace{-1ex}\\Pressure and magnetic-field effects on a quasi-2D spin-$\nicefrac{1}{2}$ Heisenberg antiferromagnet}

\author{N.\,Barbero}
\affiliation{Laboratorium f\"ur Festk\"orperphysik, ETH Z\"urich, CH-8093 Zurich, Switzerland}

\author{T.\,Shiroka}\email[Corresponding author: \vspace{8pt}]{tshiroka@phys.ethz.ch}
\affiliation{Laboratorium f\"ur Festk\"orperphysik, ETH Z\"urich, CH-8093 Zurich, Switzerland}
\affiliation{Paul Scherrer Institut, CH-5232 Villigen PSI, Switzerland}

\author{C. P. \,Landee}
\affiliation{Department of Physics, Clark University, Worcester, Massachusetts 01610, USA}

\author{M.\,Pikulski}
\affiliation{Laboratorium f\"ur Festk\"orperphysik, ETH Z\"urich, CH-8093 Zurich, Switzerland}

\author{H.-R.\,Ott}
\affiliation{Laboratorium f\"ur Festk\"orperphysik, ETH Z\"urich, CH-8093 Zurich, Switzerland}
\affiliation{Paul Scherrer Institut, CH-5232 Villigen PSI, Switzerland}

\author{J.\,Mesot}
\affiliation{Laboratorium f\"ur Festk\"orperphysik, ETH Z\"urich, CH-8093 Zurich, Switzerland}
\affiliation{Paul Scherrer Institut, CH-5232 Villigen PSI, Switzerland}

\begin{abstract}
Cu(pz)$_2$(ClO$_4$)$_2$ (with \textit{pz} denoting pyrazine, C$_4$H$_4$N$_2$) 
is among the best realizations of a two-dimensional spin-$\nicefrac{1}{2}$ 
square-lattice antiferromagnet. 
Below $T_{\mathrm{N}}$ = 4.21\,K, its weak interlayer couplings induce a 3D magnetic 
order, strongly influenced by external magnetic fields and/or hydrostatic pressure. 
Previous work, focusing on the [$H,T$] phase diagram, identified a spin-flop transition, 
resulting in a field-tunable bicritical point. However, the influence of external pressure 
has not been investigated yet. 
Here we explore the extended [$p,H,T$] phase diagram of Cu(pz)$_2$(ClO$_4$)$_2$ 
under pressures up to 12\,kbar and magnetic fields up to 7.1\,T, via magnetometry 
and $^{35}$Cl nuclear magnetic resonance (NMR) measurements. 
The application of magnetic fields enhances $T_{XY}$, the crossover temperature from 
the Heisenberg to the $XY$ model, thus pointing to an enhancement of the effective 
anisotropy. The applied pressure has an opposite effect 
[$\mathrm{d}T_\mathrm{N}/\mathrm{d}p = -0.050(8)$\,K/kbar], as it modifies marginally 
the interlayer couplings, but likely changes more significantly the orbital reorientation 
and the square-lattice deformation. This results in a remodeling of the effective 
Hamiltonian, whereby the field and pressure effects compensate each other. 
Finally, by comparing the experimental data with numerical simulations we estimate 
$T_{\mathrm{BKT}}$, the temperature of the Berezinskii-Kosterlitz-Thouless topological transition 
and argue why it is inaccessible in our case.
\end{abstract}


%


\maketitle\enlargethispage{3pt}

\vspace{-5pt}
\section{Introduction}\enlargethispage{8pt}
The antiferromagnetic coupling of magnetic moments in materials with 
low-dimensional sub-units gives rise to interesting phenomena, including 
the Haldane gap, the quantum spin-liquid state, topological phase 
transitions, etc.\cite{Rice2001} 
Such systems comprise both one- (chains and ladders), as well as two-dimensional  
(triangular-, square-, and rectangular) spin arrangements, which can be 
realized by synthesizing the relevant materials. These can serve as 
playgrounds for testing key theoretical concepts, such as those of quantum 
phase transitions and criticality.\cite{Sachdev2008} 
In addition, studies of quantum magnets and spin-wave dynamics promise 
new applications in the fields of magnonics and spintronics,\cite{Di2015} 
where the use of antiferromagnets (AFMs) may serve to improve 
magnetic-storage capabilities.\cite{Soh2011}  

The two-dimensional Heisenberg-type cuprates deserve special attention, 
since their weakly-interacting layers, characterized by a square-lattice 
disposition of Cu(II) ions, display remarkable structural and magnetic 
analogies with the copper-oxide high-$T_c$ superconductors, such as, e.g., 
La$_{2-x}$Sr$_x$CuO$_4$.
In this work we focus on Cu(pz)$_2$(ClO$_4$)$_2$, a prototype material 
with properties that are governed by the competition between order and 
quantum fluctuations, normally encountered in materials featuring 
well-isolated layers.\cite{Landee2013} Indeed, as a 2D quasi-Heisenberg 
antiferromagnet (QHAF), Cu(pz)$_2$(ClO$_4$)$_2$ exhibits quantum phase 
transitions, accessible by varying external parameters,\cite{Landau1980} 
such as applied magnetic fields or the hydrostatic pressure.

\begin{figure}[tbh]
  \centering
  \includegraphics[width=0.35\textwidth]{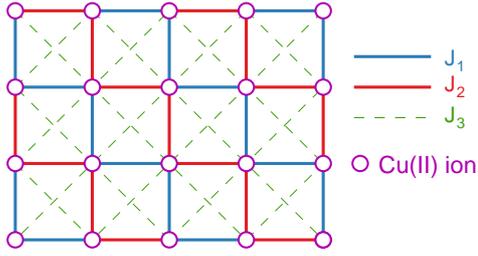} 
  \caption{\label{fig:Cupz_magnetic_structure} A layer of the 3D \CUPZ\ structure, 
showing the Cu(II) ions (circles) arranged on a square lattice. The nearest-neighbor 
exchange interactions, forming a stripy succession of $J_1$ and $J_2$ (full lines), 
are the dominant ones. The diagonal NNN $J_3$ interactions (dashed lines) are weaker.}
\end{figure}

Several experimental techniques have been used to investigate the magnetic 
properties of Cu(pz)$_2$(ClO$_4$)$_2$ in magnetic fields ranging from 0 to 
13.5\,T, including ze\-ro\--field muon-spin resonance ($\mu$SR),\cite{Lancaster2007} 
specific-heat,\cite{Lancaster2007,Tsyrulin2010} neutron scattering,\cite{Tsyrulin2010} 
ESR,\cite{Povarov2013} and magnetization\cite{Povarov2013} measurements.
The essential structural entities of \CUPZ\ consist of layers of Cu(II) 
ions with spin  $S=\nicefrac{1}{2}$, arranged on a square lattice (see 
Fig.~\ref{fig:Cupz_magnetic_structure}). The neighboring layers are bridged 
by four pyrazine ligands and their stacking involves a (\nicefrac{1}{2}, \nicefrac{1}{2}) 
unit-cell shift along the square diagonal.\cite{Darriet1979} 
The perchlorate counterions are semicoordinate in the axial sites, such that 
the Cu(II)-ions adopt a nearly tetragonal arrangement, with a clear separation 
between the layers.\cite{Landee2013} 
In \CUPZ\ the intra-layer interactions are dominated by the nearest-neighbor 
(NN) interaction among Cu(II) ions, initially estimated as 
$J/k_{\mathrm{B}} = 18.1$\,K from dispersion curves of inelastic neutron 
scattering.\cite{Tsyrulin2010} 
Subsequently, Density Functional Theory (DFT) simulations showed that a tiny 
difference in the Cu--Cu bond lengths ($d_1 = 6.898$\,\AA\ vs.\ $d_2 = 6.876$\,\AA) 
is reflected in two distinct NN exchange couplings, with $J_1/J_2 = 1.4$.\cite{Vela2013} 
By taking into account these different NN interaction strengths, the 
magnetic-interaction pattern of \CUPZ\ can be visualized as a succession of 
zigzag interaction paths (see Fig.~\ref{fig:Cupz_magnetic_structure}). 
In addition, the same simulations predict a rather weak next-nearest 
neighbor (NNN) interaction, $J_3/k_{\mathrm{B}} = 0.4$\,K, implying a 
weak degree of frustration in the system. Concerning the interlayer couplings, the 
excellent layer separation is confirmed by the low value of the $J_{\mathrm{int}}$ 
interaction ($J_{\mathrm{int}}/J = 8.8 \cdot 10^{-4}$).\cite{Tsyrulin2010} 
Nevertheless, this residual interaction is strong enough to induce a transition 
from a 2D quasi-Heisenberg AFM to a 3D ordered N\'eel state at a relatively high 
value of $T_\mathrm{N} = 4.21$\,K.\footnote{The considerable $T_\mathrm{N}$ 
values even for modest or negligible inter-to-intralayer coupling ratios, 
have a semi-empirical explanation from Monte Carlo simulations, which indicate 
an almost constant $T_\mathrm{N}$ behavior even at small $J'$ values [see 
Eq.~(\ref{eq:Neel})].\cite{Yasuda2005}
A suitable framework is provided by the self-consistent spin-wave theory 
of layered magnets with a weak interlayer coupling and magnetic 
anisotropy.\cite{Irkhin1999,Tsyrulin2010}}

In this work we primarily aimed at exploring the [$p$,$T$] phase diagram of 
Cu(pz)$_2$(ClO$_4$)$_2$, an investigation complementary to the [$H$,$T$] 
mapping of the same compound reported recently.\cite{Povarov2013} The presence 
of organic units in Cu(pz)$_2$(ClO$_4$)$_2$, generally easily deformable under 
pressure, makes it a suitable candidate for studying the role of pressure and its
influence on the strength of the different exchange interactions 
in an almost ideal 2D quantum-spin system. In addition, we aimed also at 
investigating the combined influence of external pressure and magnetic 
fields, as well as at exploring the possibility of observing the renowned 
Berezinskii-Kosterlitz-Thouless (BKT) topological 
transition\cite{Berezinskii1971,Kosterlitz1973} in the low-$T$ $XY$ phase.
In the following, we present the results of magnetometry 
(Sec.~\ref{ssec:magnetometry}) and magnetic-resonance experiments 
(Sec.~\ref{ssec:nmr_meas}), and discuss how the magnetic anisotropy, inter-, 
and intra-layer couplings are modified under the joint application of 
pressure and magnetic field. 
The unexpected compensation effects we observe, seem to arise from the 
contrasting roles played by the deformation of the 2D square lattice 
and the reorientation of the electronic orbitals of in-plane ions.

\vspace{-5pt}
\section{Experimental Details\label{sec:Experimental_Details}}
Single crystals of Cu(pz)$_2$(ClO$_4$)$_2$ were synthesized by dissolving 
Cu(ClO$_4$)$_2\cdot$6H$_2$O and pyrazine in water to which a drop of dilute 
HClO$_4$(aq) was added [the latter prevents the precipitation of Cu(OH)$_2$ 
and CuCO$_3$].\cite{Woodward2007} The solution was then left to evaporate, 
with the crystals growing on a timescale of weeks. Eventually the mixture 
was filtered and the recovered crystals washed in cold deionized water and 
left to dry in air. Since the samples are air- and water sensitive, they were 
constantly kept in a dessicator.\cite{Woodward2007} 
Three samples with different sizes were available: one large sample (81\,mg, 
6.5$\times$5$\times$1 mm$^3$) was used for the zero-pressure magnetometry, 
whereas the smaller ones (14 and 5.5\,mg) were employed for the NMR and 
magnetization measurements under pressure. 
The single Cu(pz)$_2$(ClO$_4$)$_2$ crystals were cut such as to have 
two opposite faces parallel to the $bc$ planes. This allowed us to align the 
samples with their $a$ axis (\textit{hard} magnetic axis) parallel to the applied 
magnetic field with an uncertainty of a few degrees.

The SQUID magnetometry measurements at zero pressure were carried out using 
a commercial Magnetic Property Measurement System (MPMS) setup from Quantum Design, equipped with a 5.5\,T-magnet.
For the NMR investigations of Cu(pz)$_2$(ClO$_4$)$_2$, i.e., line- and spin-lattice 
relaxation measurements at 7.063\,T, the $^{35}$Cl nucleus proved to be the 
most suitable for our study. 
Typically, we employed standard spin-echo sequences, 
consisting in $\nicefrac{\pi}{2}$ and $\pi$ pulses of 2 and 4\,$\mu$s, respectively, 
with a recycle delay ranging from 0.2 to 0.6\,s, depending on temperature. 
The narrow lines at low pressure were obtained by the FFT of the echo signal, 
while the broader high-pressure lines were acquired by sweeping the frequency 
and summing the resulting FFT data.
A limited number of scans (from 256 to 1024) was sufficient for acquiring signals 
with a good S/N ratio. The $^{35}$Cl NMR signal of an aqueous NaCl solution at 
29.467\,MHz (measured separately at room temperature) served as a frequency 
reference.
\begin{figure}[tbh]
  \centering
  \includegraphics[width=0.42\textwidth]{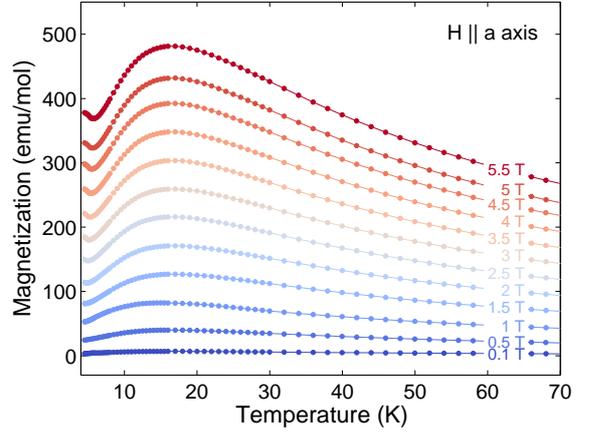} 
  \caption{\label{fig:2_SQUID_magn_mol_field} Magnetization $M(T)$ curves collected 
  at different magnetic fields (from 0.1 to 5.5\,T), applied along the $a$ axis.}
\end{figure}

For the hydrostatic pressure studies, two high-pressure cells were built, 
with bodies and nuts consisting of high-purity, low-magnetic-response BeCu 
and bores of 5 and 2.7 mm for the NMR and SQUID-magnetometry measurements, 
respectively. The feedthrough assembly of the NMR pressure cell was constructed 
using a hardened MP35N alloy, while the piston consisted of tungsten carbide (WC).\cite{Eremets1996} 
The leads were insulated copper wires with a diameter of 0.1\,mm. 
Once inserted in the NMR coil, the specimen was embedded into a PTFE container 
filled with silicon oil. In addition, a secondary coil, containing a powdered, 
high-purity Cu(I)-oxide sample mixed with paraffin wax and serving as an in-situ 
pressure gauge, was also inserted into the container, which was then sealed with 
the feedthrough unit.
The pressurization was performed by means of an industrial press (25-T Enerpac), 
with which pressures up to 18\,kbar were achieved at room temperature (decreasing 
by ca.\ 25\% at low temperatures). The pressure gauge was calibrated via $^{63}$Cu 
NQR measurements, following established references.\cite{Reyes1992} 
The $^{63}$Cu NQR signal in Cu$_2$O was acquired using a spin-echo sequence 
with a recycle delay of 60\,s. 
The pressure cell used in the SQUID magnetometer had a similar piston-design 
as the NMR cell. However, due to magnetometer-related dimensional constraints, 
only pressures up to about 12\,kbar could be reached in this case. In the 
high-pressure region of the cell, the sample and the pressure gauge, in the 
form of a 1-mg piece of lead, were separated by a quartz-rod spacer. The in-situ 
pressure was monitored by measuring the superconducting transition of lead, by 
considering its known $T_c$-vs.-pressure dependence.\cite{Clark1978}

\vspace{-5pt}
\section{Experimental results and discussion\label{sec:Results}}
\subsection{SQUID magnetometry\label{ssec:magnetometry}}
\subsubsection{\label{ssec:nmagnetizationMTH}Magnetic field effects on magnetization}
\begin{figure}[tbh]
  \centering
  \includegraphics[width=0.42\textwidth]{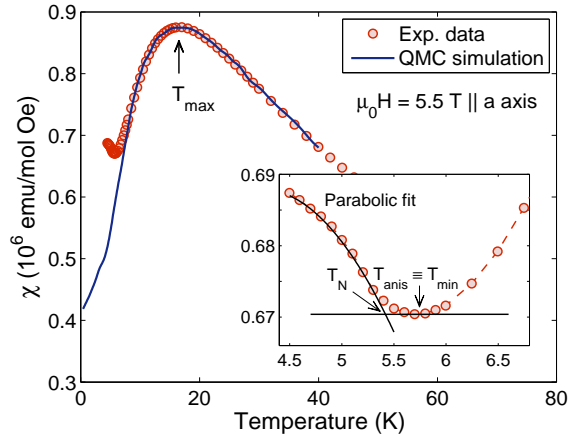} 
  \caption[Magnetic susceptibility data vs.\ QMC 
simulation]{\label{fig:SQUID_simulation}Comparison of the temperature-dependent 
magnetic susceptibility (dots) with Quantum Monte Carlo (QMC) simulations of a 
uniform susceptibility model (line). The simulations were carried out on an 
$L\times L$ lattice, with $L = 256$ sites, based on the ALPS package using the 
Stochastic Series Expansion (SSE) method. Inset: Close-up view of the magnetic 
susceptibility curve, showing the definitions of $T_{\mathrm{N}}$ and $T_{\mathrm{min}}$.}
\end{figure}

In Fig.~\ref{fig:2_SQUID_magn_mol_field} we display the results of measurements 
of the temperature dependence of the magnetization $M(T)$ of \CUPZ\ between 
4.5\,K and room temperature at different external magnetic fields, from 0.1\,T 
up to 5.5\,T. From these data we extracted three distinct temperatures: 
$T_{\mathrm{max}}$, $T_{\mathrm{min}}$, and $T_{\mathrm{N}}$, as indicated 
in Fig.~\ref{fig:SQUID_simulation}.

The values of $T_{\mathrm{max}}$, determined from parabolic fits to the 
data around the magnetization maximum, can be related with the results 
of high-temperature series expansion calculations,\cite{Lines1970} 
or to those of more recent numerical simulations,\cite{Kim1998} 
which for a 2D Heisenberg spin-$S$ system predict 
$k_{\mathrm{B}} T_{\mathrm{max}}/J = 1.12\,S (S+1) + 0.10$ and 
$k_{\mathrm{B}} T_{\mathrm{max}}/J = 0.936$, respectively.
By considering the average experimental value $T_{\mathrm{max}}\approx 17$\,K, 
we obtain an average value for the in-plane exchange-interaction parameter 
$J/k_{\mathrm{B}} = 18.2$\,K, in excellent agreement with the above-mentioned 
$J/k_{\mathrm{B}} = 18.1$\,K, derived from neutron-scattering data.\cite{Tsyrulin2010} 
This suggests that, at temperatures above 10\,K, the system may be modeled 
as an isotropic two-dimensional Heisenberg antiferromagnet (HAFM). In order 
to check this assumption we performed Quantum Monte Carlo (QMC) simulations 
on a $L \times L$ lattice, with $L=256$ sites, based on the ALPS package\cite{Bauer2011}  and 
using the above mentioned $J$ as the only parameter. As shown in Fig.~\ref{fig:SQUID_simulation},
for $T>10$\,K the calculated $\chi(T)$ dependence is in very good agreement with
experimental susceptibility data, thus confirming the above-mentioned 2D-model assumption, 
while the simulation clearly fails to reproduce the data at lower temperatures. 
In two-dimensional spin-$\nicefrac{1}{2}$ HAFMs, a minimum in $\chi(T)$ is 
claimed to indicate in-plane exchange anisotropies $\Delta_{XY}$ which, even 
when tiny, may induce a crossover from a Heisenberg to an $XY$ regime.\cite{Ding1992}

As we shall see below, the 3D ordering temperature $T_{\mathrm{N}}$ may 
accurately be determined by measurements of the NMR spin-lattice relaxation 
rate ${T_1}^{-1}(T)$. The evaluation of $T_{\mathrm{N}}$ from the $\chi(T)$ data, 
however, is not straightforward. In a previous work,\cite{Xiao2009} it was claimed 
that the inversion point in $\chi(T)$ below $T_{\mathrm{min}}$ provides a 
reliable estimate of $T_{\mathrm{N}}$. For our purposes, mainly to monitor 
the variation of  $T_{\mathrm{N}}$ upon changing the magnitude of the external 
magnetic field, we have chosen the approach indicated in the inset of 
Fig.~\ref{fig:SQUID_simulation}. Here we estimate $T_{\mathrm{N}}$ by a 
parabolic fit to the data at the lower end of the covered temperature range 
and its extrapolation to the minimum value in $\chi(T)$.

Following the approach outlined in earlier work on Sr$_2$CuO$_2$Cl$_2$,\cite{Cuccoli2003} 
another model compound for 2D magnetic systems, we distinguish three regions 
in the low-temperature range of $\chi(T)$, i.e., $T< T_{\mathrm{N}}$, 
$T_{\mathrm{N}} < T < T_{\mathrm{min}}$, and $T > T_{\mathrm{min}}$. 
Above approximately 10\,K, i.e., distinctly above $T_{\mathrm{min}}$, 
\CUPZ\ may be regarded as a 2D $S = \nicefrac{1}{2}$ HAFM. In a narrow 
temperature range just below $T_{\mathrm{min}}$, the system adopts an 
intermediate disordered $XY$ phase, which undergoes a 3D ordering at 
$T_{\mathrm{N}}$. It was empirically established that the crossover 
temperature $T_{XY}$, that we tentatively set equal with $T_{\mathrm{min}}$, 
may be estimated by:\cite{Cuccoli2003} 
\begin{equation} \label{eq:Cuccoli}
T_{XY} = \frac{4\pi \cdot 0.214 \cdot J}{\ln(160/\Delta_{XY})},
\end{equation}
where $\Delta_{XY}$ is the intrinsic in-plane exchange anisotropy. Eq.~\ref{eq:Cuccoli}
implies that $T_{XY}$ varies linearly with $J$, but tends to saturate 
at small $\Delta_{XY}$ values. By inserting the values of 
$T_{XY}\approx T_{\mathrm{min}}$ we obtain values for $\Delta_{XY}$ of the 
order of $10^{-3}$ to $10^{-2}$. More specifically, we note that an increase 
of the magnetic field enhances the resulting value of $\Delta_{XY}$, 
which rises from $6\times10^{-3}$ at 2\,T to $2.4\times10^{-2}$ at 4\,T. 
Given the tiny in-plane exchange anisotropy $\Delta_{XY}$ of \CUPZ\ (see above), 
it is of interest to investigate its low-temperature behavior in detail. 
\begin{figure}[tbh]
  \centering
  \includegraphics[width=0.4\textwidth]{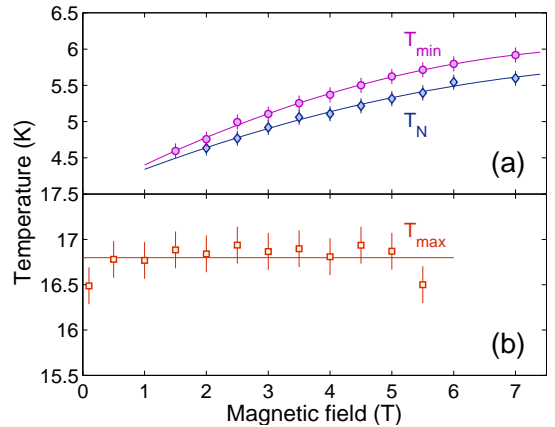} 
  \caption{\label{fig:4_SQUID_temp_field} Metadata from Fig.~\ref{fig:2_SQUID_magn_mol_field} 
showing (a) the evolution of $T_{\mathrm{min}}$, $T_{\mathrm{N}}$, and of (b) $T_{\mathrm{max}}$ with the applied magnetic field at $p = 0$\, kbar. While $T_{\mathrm{min}}$ and $T_{\mathrm{N}}$  gradually increase with the field, $T_{\mathrm{max}}$ remains practically constant.}
\end{figure}

In order to capture the influence of external magnetic fields on the low-temperature 
magnetic features of \CUPZ, one can use a general Hamiltonian $\mathcal{H}$, which 
includes an applied magnetic field $H$, NN and NNN interactions ($J_{\mathrm{NN}}$ 
and $J_{\mathrm{NNN}}$, respectively), and a generic exchange anisotropy parameter 
$\Delta_{XY}$:\cite{Tsyrulin2010} 
\begin{eqnarray}
\label{eq:Hamiltonian}
\mathcal{H} =&& J_{\mathrm{NN}}\sum_{<i,j>}^{} [{S_i}^x{S_j}^x+{S_i}^y{S_j}^y+(1- \Delta_{XY}) {S_i}^z{S_j}^z]\nonumber\\
&&+ J_{\mathrm{NNN}}\sum_{<i,k>}^{} {\mathbf{S}_i} \cdot {\mathbf{S}_k} - g \mu_{\mathrm{B}}H\sum_{i}^{} \mathbf{S}_i.
\end{eqnarray}
For $\Delta_{XY} = 0$  Eq.~(\ref{eq:Hamiltonian}) reproduces the Heisenberg 
case, while for $\Delta_{XY} = 1$ it depicts the $XY$ model. Earlier numerical 
simulation studies of the two-dimensional $XY$ model predicted a significant 
upturn in the magnetic susceptibility when $T_{\mathrm{BKT}}$ is 
approached.\cite{Gupta1988}
This behavior was observed experimentally in Sr$_2$CuO$_2$Cl$_2$,\cite{Cuccoli2003} 
where the susceptibility data and numerical simulations were found to be in good 
agreement. A short discussion on the observability of a BKT transition in our 
case is given in Sec.~\ref{sec:BKT_discussion}.
From the data shown in Fig.~\ref{fig:2_SQUID_magn_mol_field} it may be 
inferred that the crossover from the Heisenberg to $XY$ behavior and the 
transition to a 3D magnetic order both shift to higher temperatures with 
increasing magnetic field (see Fig.~\ref{fig:4_SQUID_temp_field}).
The main cause for the first is most likely a 
growing anisotropy with increasing field, while for the second a field-induced 
reduction of quantum fluctuations. This conclusion reflects the fact that 
higher magnetic fields enhance the trend for order, quench the quantum 
fluctuations, and enhance the anisotropy, such that the field orientation 
imposes the energetically-favorable direction. As expected, the value of 
$J$, related to $T_{\mathrm{max}}$, is essentially field independent.

\begin{figure}[tbh]
  \centering
  \includegraphics[width=0.42\textwidth]{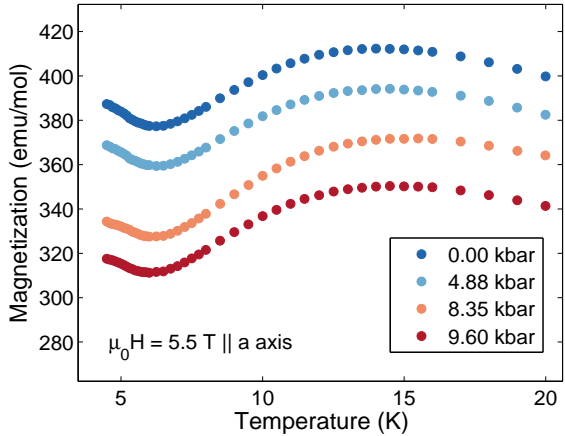} 
  \caption{\label{fig:SQUID_MTP} Magnetization $M$ vs.\ temperature for 
different applied pressures (from 0 to 9.60\,kbar). Data were acquired 
in an external magnetic field of 5.5\,T, parallel to the $a$ axis 
and subsequently corrected for a small contribution to $M(T)$ 
of the empty pressure-cell body.}
\end{figure}

\subsubsection{Influence of external pressure}
\begin{figure}[tbh]
  \centering
  \includegraphics[width=0.42\textwidth]{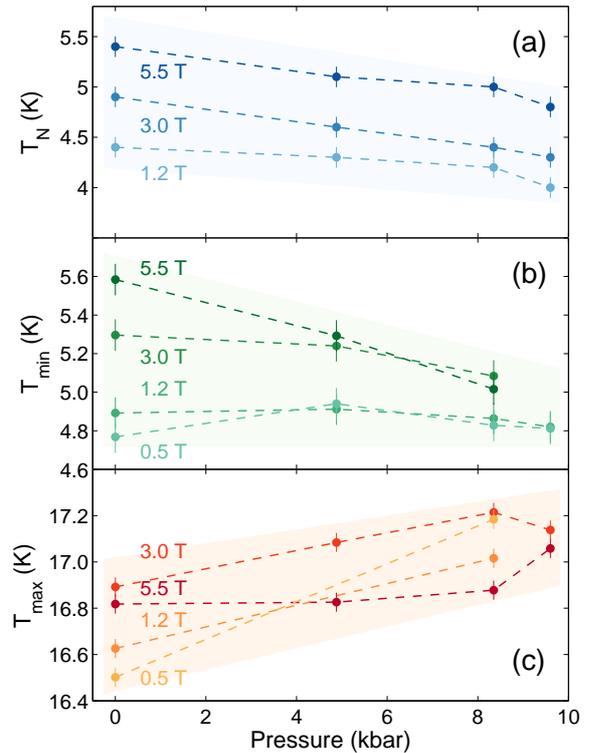} 
  \caption{\label{fig:SQUID_chi_press} $T_{\mathrm{N}}$ (a), 
$T_{\mathrm{min}}$ (b), and $T_{\mathrm{max}}$ (c) vs.\ applied pressure 
for a selection of magnetic field values. While the first two show a 
decreasing trend with pressure, $T_{\mathrm{max}}$ exhibits an opposite 
behavior (shaded areas). A compensation of pressure and magnetic-field 
effects (especially at higher fields) prevented a more detailed analysis.}
\end{figure}

As an example of the influence of external hydrostatic pressure at a fixed 
magnetic field, we show a corresponding $M(T)$ data set at selected pressures 
in a field of 5.5\,T in Fig.~\ref{fig:SQUID_MTP}.
On the basis of such data, we evaluated the pressure dependence of 
$T_{\mathrm{N}}$, $T_{\mathrm{min}}$, and $T_{\mathrm{max}}$ for a selection of 
external fields; the overall results are displayed in Fig.~\ref{fig:SQUID_chi_press}.

As we mentioned above, an exact determination of $T_{\mathrm{N}}$ from 
$M(T)$ data is not straightforward. Nevertheless, an acceptable way to 
estimate $T_{\mathrm{N}}$ is indicated in the inset of Fig.~\ref{fig:SQUID_simulation}. 
Following this procedure, we obtain a clear trend of $T_{\mathrm{N}}(p)$, which, 
in a field of 5.5\,T, decreases monotonously from 5.4\,K at $p$ = 0 to 4.8\,K 
at $p$ = 9.6\,kbar [see Fig.~\ref{fig:SQUID_chi_press}(a)].

Similarly, from panel (b) of this figure it may be seen that at low fields 
also $T_{\mathrm{min}} \approx T_{XY}$ decreases with increasing pressure. 
At higher fields, the pressure-induced shifts of $T_{\mathrm{min}}$ are 
clearly reduced in magnitude with no distinct trend. Quite generally, the 
effect of external pressure on the $M(T)$ curve of \CUPZ\ is rather weak. 
Nevertheless, the trends in $T_{\mathrm{min}}(p)$ and $T_{\mathrm{max}}(p)$ 
observed in low external fields seem to be weakened in elevated fields.

In a first approach, to interpret the observed pressure effects on 
$T_\mathrm{N}$, one can use the Ehrenfest relation for second-order 
phase transitions:\cite{Pippard1966}
\begin{equation} 
\label{eq:Ehrenfest}
\frac{\mathrm{d}T_{\mathrm{N}}}{\mathrm{d}p} = \frac{T V_m \Delta \beta}{\Delta C_p}.
\end{equation}
Here $p$ is the pressure, $V_m$ the molar volume, $C_p$ the specific heat 
at constant pressure, and $\beta$ the volumetric coefficient of thermal 
expansion. The quantities $\Delta\beta$ and $\Delta C_p$ denote the changes 
of the volume thermal-expansion coefficient and of the specific heat at 
constant pressure across the transition to 3D AFM order. 

In Eq.~\ref{eq:Ehrenfest} the product $TV_m$ is always positive, while from specific-heat 
measurements it is known that $\Delta C_p > 0$ for \CUPZ.\cite{Tsyrulin2010} 
A negative value of $\mathrm{d}T_{\mathrm{N}}/\mathrm{d}p$ requires 
that $\Delta\beta$ is negative, but this has not been confirmed by any experiment yet. 
At any rate, as we pointed out above, the results of NMR experiments to be described below, 
confirm that $T_{\mathrm{N}}$ decreases monotonously with increasing pressure.

The occurrence of a negative $\Delta\beta$ and hence $\mathrm{d}T_{c}/\mathrm{d}p < 0$ 
across a phase-boundary is quite common for conventional superconductors 
and is usually interpreted as being due to an enhanced stiffness of the crystal 
lattice at higher pressure. In our case, $T_{\mathrm{N}}$ depends primarily on 
the strength of the intraplane ($J$) and interplane ($J_{\mathrm{int}}$) exchange 
interactions between the Cu ions. For $J_{\mathrm{int}}$ = 0, i.e., the ideal 
2D Heisenberg AFM situation, a transition to an ordered state is not expected 
for $T > 0$.\cite{Mermin1966} 
For $J_{\mathrm{int}}\neq$ 0, it was concluded from QMC simulations that 
the relation between $T_{\mathrm{N}}$ and $J_{\mathrm{int}}/J$ can be 
described quite accurately by:\cite{Yasuda2005}
\begin{equation} \label{eq:Neel}
T_\mathrm{N} = \frac{4 \pi \rho_S}{b-\ln{(J_\mathrm{{int}}/J)}}.
\end{equation}
Here $\rho_S = 0.138J$ is the spin-stiffness parameter and $b = 2.43$ for 
$S = \nicefrac{1}{2}$. Thus the observed reduction of $T_{\mathrm{N}}$ under 
pressure requires that either $J$ or the ratio $J_{\mathrm{int}}/J$ decreases. 
The second requirement can be fulfilled if $J$ increases with pressure, in 
contrast with the first requirement or, if $J_{\mathrm{int}}$ is reduced 
substantially, hardly to be expected upon increasing pressure. According to the 
data shown in Fig.~\ref{fig:SQUID_chi_press}(b), $T_{\mathrm{max}}(p)$ and, 
therefore, $J(p)$ exhibit a trend of enhancement with increasing pressure. 
Taken together, these observations imply that on the basis of the presently 
available data, there is no straightforward explanation for the 
$\mathrm{d}T_\mathrm{N}/\mathrm{d}p < 0$ behavior.

\subsection{NMR measurements\label{ssec:nmr_meas}}
In what follows we describe a ${}^{35}$Cl NMR study aimed at exploring the 
low-temperature magnetic order in \CUPZ. For this purpose a single-crystal 
specimen was mounted on the sample-holder and aligned with its hard axis $a$ 
parallel to the externally applied static field $\mu_{\mathrm{B}}H$ = 7.066\,T.
\begin{figure}[tbh]
  \centering
  \includegraphics[width=0.42\textwidth]{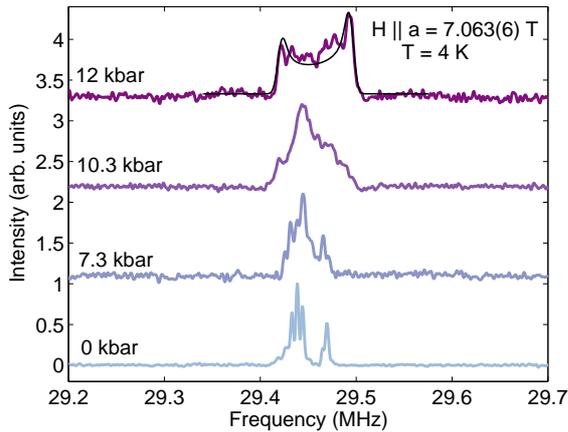} 
  \caption{\label{fig:8_NMR_line_LT_press_compar}Low-temperature $^{35}$Cl 
NMR lineshapes at four different pressures. 
The progressive line broadening with pressure suggests the development 
of an incommensurate magnetic structure at 12\,kbar. The black line represents 
a model calculation (see text).}
\end{figure}

From previous investigations it is known that in a field of approximately 
0.5\,T, \CUPZ\ exhibits a spin-flop transition below $T_{\mathrm{N}}$, 
resulting in a bicritical point in the ($H$,$T$) phase diagram.\cite{Povarov2013} 
In such a low magnetic field, the S/N ratio of the NMR experiments, 
however, does not allow for accumulating reliable data in a reasonable 
amount of time and, therefore, this transition cannot be monitored via 
$^{35}$Cl NMR. 

The onset of the AFM order in zero field was first established at 
$T_{\mathrm{N}}$ = 4.21\,K via $\mu$SR experiments.\cite{Lancaster2007} 
Later on it was reported that the boundary fixing $T_{\mathrm{N}}$ extends 
up to higher fields of at least 12\,T with $\mathrm{d}T_{\mathrm{N}}/\mathrm{d}H > 0$.\cite{Tsyrulin2010} 
For our measurements we have chosen an external field of approximately 7\,T (see above). 
By inserting the specimen into a high-pressure cell, we were able to monitor the pressure 
effects on $T_{\mathrm{N}}$, from ambient pressure up to $p = 12$\,kbar.

In Fig.~\ref{fig:8_NMR_line_LT_press_compar} we display a selection of 
${}^{35}$Cl NMR lines measured at\,4 K, i.e., below $T_{\mathrm{N}}$, at 
four different externally applied hydrostatic pressures of 0.0, 7.3, 10.3, and 
12\,kbar. The NMR lineshapes reflect the local distribution of magnetic fields 
and hence provide information on the magnetic structure of the ordered phase. 
Due to two magnetically inequivalent sites, at zero pressure the spectra feature 
two main peaks, similar to those observed above $T_{\mathrm{N}}$ (not shown), 
where the absence of quadrupole splitting reflects the high (tetrahedral) symmetry 
of the perchlorate (ClO$^{4-}$) counterions. A lowering of the symmetry  
below $T_{\mathrm{N}}$, gives rise to a tiny (5.6\,kHz) quadrupole splitting 
in at least one of the sites which, by taking into account the ${}^{35}$Cl quadrupole 
moment, $Q = -8.2\times10^{-30}$\,m$^2$, implies an electric field gradient 
$eq \sim 2\times10^{19}\,$V/m$^2$. 
The separate peaks observed at zero applied pressure gradually  broaden and 
mix upon increasing pressure. The broadening of the lines and the growing 
irregularity of their shapes indicate a growing complexity of the local magnetization. 
It is conceivable that the soft ClO$_4$ ligands bridging the Cu-ion planes cause 
internal deformations of the unit cells, thereby enhancing the complexity of the 
two AFM sublattices and lead to a commensurate-to-incommensurate crossover. 
Indeed, the lineshape at 12\,kbar is reminiscent of an incommensurate magnetic 
structure. A simplified model (see, e.g., Ref.~\onlinecite{Blinc1981,Gippius2008}) 
with a magnetic splitting $\Delta_M = 69$\,kHz (0.165 T) reproduces the 
experimental data reasonably well. In addition, the asymmetric line shape indicates 
a bunching of the Cu$^{2+}$ magnetic moments along the applied field direction.

\begin{figure}[tbh]
  \centering
  \includegraphics[width=0.42\textwidth]{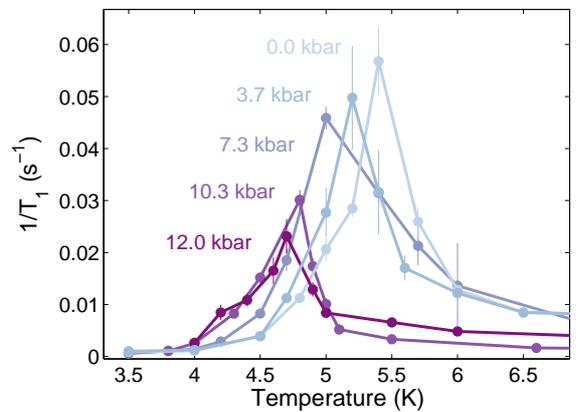} 
  \caption{\label{fig:6_NMR_LT_pressure_compar}The $1/T_1$ vs.\ $T$ dependence 
shows clear peaks at $T_\mathrm{N}$. As the pressure increases the peaks become 
less prominent (slower spin-lattice relaxation times) and shift toward lower 
temperatures. Lines are guides to the eye.}
\end{figure}

A study of the Cu(II)-based coordination network compound CuF$_2$(H$_2$O)$_2$(pz) 
has already shown that hydrostatic pressure can drastically change the 
magnetic properties of this class of materials.\cite{Halder2011} In particular, 
since some bonds and bridges are \textit{softer} and others more \textit{rigid}, 
pressure can induce non-isotropic deformations, including both axial and angular 
distortions. Similarly, in the \CUPZ\ case, the interlayer interactions and 
the perchlorate gegenanions are expected to be more sensitive to distortions 
than the pyrazine bridges. This would imply an increase in $T_\mathrm{N}$ 
with pressure, reflecting a reduced interlayer distance, an increased number 
of nearest neighbors and, as a consequence, a reduction of quantum fluctuations.
Experimental data, as those shown in Fig.~\ref{fig:6_NMR_LT_pressure_compar}, 
however, clearly contradict this hypothesis. 

\begin{figure}[tbh]
  \centering
  \includegraphics[width=0.42\textwidth]{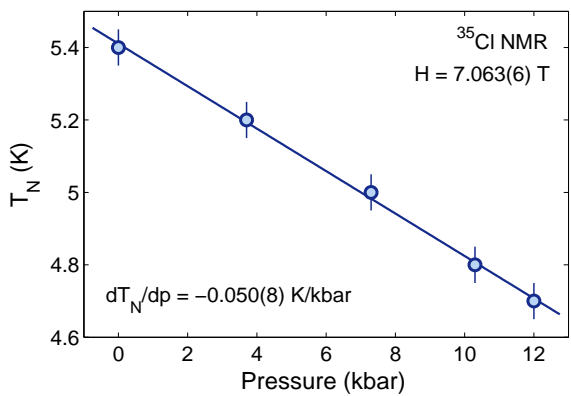} 
  \caption{\label{fig:7_NMR_TN_press_compar}$T_\mathrm{N}$ dependence 
on applied pressure exhibits a decreasing trend with a 
$\mathrm{d}T_\mathrm{N}/\mathrm{d}p = -0.050(8)$\,K/kbar slope. The 
linear-fit approximation is acceptable within this range of pressures.}
\end{figure}

The NMR spin-lattice relaxation rate $T_1(T)$ is expected to exhibit a peak 
at the transition to magnetic order.\cite{Borsa2007} This is clearly confirmed 
in our case by the data displayed in Fig.~\ref{fig:6_NMR_LT_pressure_compar}, 
which show $1/T_1(T)$  for different applied pressures. The covered temperature 
range from 3.5 to 8\,K was sufficient to adequately explore the accessible phase 
boundary. The spin-lattice relaxation times $T_1$ were obtained by means of the 
inversion-recovery method. The magnetization recovery was fitted using:\cite{Mcdowell1995}
\begin{equation}
\label{eq:T1_relax}
M_z(t)/M_0 = 1 - f\,[0.9\cdot e^{-(6t/T_1)^{\lambda}}+0.1 \cdot e^{-(t/T_1)^{\lambda}}], 
\end{equation}
an equation valid for spin-$\nicefrac{3}{2}$ nuclei. Here $M_0$ is the 
saturation value of the nuclear magnetization, $f$ is an amplitude parameter 
(ideally 2 for a complete inversion), while the stretching coefficient 
$\lambda$ accounts for the distribution of the spin-lattice relaxation times 
around a characteristic value $T_1$. Usually $\lambda$ = 1 at high temperatures 
indicates a well-defined single spin-lattice relaxation rate $1/T_1$. At lower 
temperatures, $\lambda$ often starts to decrease, indicating a growing 
distribution of $1/T_1$ values.
In our case $\lambda(T)$ is of the order of 1 at the selected pressures 
across the covered temperature range. We notice that the intensity of the 
peaks, corresponding to the fastest relaxation times, is less prominent as 
the pressure increases. This observation indicates that with increasing pressure, 
the phase boundary is less well defined and/or the onset of incommensurate 
magnetism occurs, a feature also suggested by the line-shapes (see above). 
The $T_{\mathrm{N}}(p)$ plot shown in Fig.~\ref{fig:7_NMR_TN_press_compar} 
reveals a linearly decreasing $T_{\mathrm{N}}$, characterized by a slope 
$\mathrm{d}T_{\mathrm{N}}/\mathrm{d}p = -0.050(8)$\,K/kbar.

Two factors may account for the observed behavior: (i) the intrinsic rhombic 
distortion of the Cu sublattice, due to a small 0.3\% difference between the 
2D lattice diagonals, causes an in-plane anisotropy,\cite{Povarov2013} and (ii) 
the anisotropic response of the system under hydrostatic pressure. Both of 
them result in angular deformations which quantitatively oppose the AFM 
ordering and shift the magnetic transition towards lower temperatures.

\vspace{-5pt}
\section{\label{sec:BKT_discussion}Accessibility of the BKT phase transition}
As we argued above, the temperature dependence of magnetization $M(T)$ in 
\CUPZ\ indicates a crossover from an isotropic Heisenberg-AFM to an $XY$-model 
behavior at low temperatures. An $XY$ phase is expected to be unstable with 
respect to a topological Berezinskii-Kosterlitz-Thouless (BKT) 
transition\cite{Berezinskii1971,Kosterlitz1973} at $T_{\mathrm{BKT}} < T_{\mathrm{XY}}$. 
Below $T_{\mathrm{BKT}}$, a quasi-long range order (QLRO) is established, 
where the spins are aligned and the correlation-function decays with a power 
law, while at $T>T_{\mathrm{BKT}}$, a disordered high-temperature phase with 
exponentially decaying correlations is adopted. Although in the QLRO phase, 
quantum fluctuations are strong enough to destroy any long-range order, when 
the alignment of a spin changes, the nearby spins can still follow the new 
alignment, hence preserving the local order. The vortices, a topologically 
stable arrangement of spins, lower the correlation of the system: at temperatures 
below the topological transition they exist in vortex-antivortex bound states 
with zero net vorticity.\cite{Sandvik2010} On the other hand, above the transition, 
a proliferation of unbound vortices occurs.
As reported in the literature, QMC simulations regarding the two-dimensional 
$XY$ model allow for an evaluation of the numerical value of $T_{\mathrm{BKT}}$ 
by\cite{Ding1992}  
\begin{equation}
\label{eq:TBKT}
T_\mathrm{BKT} = 4 \pi \rho_s (1-0.491\Delta_{XY})/\ln(419/\Delta_{XY}),
\end{equation}
with $\rho_s$ defined as in Eq.~(\ref{eq:Neel}). 
Since $T_\mathrm{BKT} \propto 1/\ln(1/\Delta_{\mathrm{XY}})$, a negligibly 
small decrease of $T_{\mathrm{BKT}}$ is expected when $\Delta_{XY}\rightarrow 0$, 
even at low  $\Delta_{XY}$ values. For instance, $k_\mathrm{B}T_{\mathrm{BKT}}$ 
decreases only marginally (from ${\sim}0.285 J$ to $0.25 J$) upon a fivefold 
reduction of $\Delta_{XY}$ (from 0.1 to 0.02). 
By inserting the value of $\Delta_{XY} = 4.6\times10^{-3}$ for 
\CUPZ\cite{Xiao2009} in Eq.~(\ref{eq:TBKT}), a relatively high value of 
$T_{\mathrm{BKT}}= 2.75$\,K is obtained. This, however, is still distinctly 
lower than $T_{\mathrm{N}}$, implying that in \CUPZ, due to the prior onset 
of the three-dimensional LRO antiferromagnetic phase, the BKT phase transition 
cannot be accessed experimentally under the present circumstances. 
Extremely well separated Cu(II)-ion layers have recently been 
identified in the compound [Cu(pz)$_2$(2-hydroxypyridine)$_2$](ClO$_4$)$_2$.\cite{Selmani2010} 
The reported intralayer coupling $J/k_{\mathrm{B}} \simeq 8$\,K and a 
correspondingly low value of the interlayer coupling $J'/J \leq 0.05$\% suggest 
reductions of both $T_{\mathrm{N}}$ and $T_{\mathrm{BKT}}$ to below 2\,K. 
According to Ref.~\onlinecite{Selmani2010}, the compound does not order magnetically 
above 1.8\,K. The possibility that in this case $T_{\mathrm{BKT}} > T_{\mathrm{N}}$ 
cannot be ruled out and should be checked experimentally.

\vspace{-5pt}
\section{Summary and conclusions\label{sec:Conclusions}}
The two-dimensional spin-$\nicefrac{1}{2}$ square-lattice antiferromagnet 
Cu(pz)$_2$(ClO$_4$)$_2$ was studied by means of SQUID magnetometry and 
${}^{35}$Cl NMR at different magnetic fields and hydrostatic pressures, 
which allowed for a mapping of the [$H$,$T$] and [$p$,$T$] phase diagrams 
up to 7\,T and 12\,kbar, respectively. We argue that the observed increase 
of the 3D ordering ($T_\mathrm{N}$) and the $XY$ crossover ($T_{\mathrm{XY}}$) 
temperatures in large applied magnetic fields is due to both an enhancement of 
the effective $XY$ anisotropy and the quenching of quantum fluctuations. 
Taken together, the data shown in Fig.~\ref{fig:4_SQUID_temp_field} and 
Fig.~\ref{fig:SQUID_chi_press} imply some degree of compensation between 
the effects of external magnetic field and external hydrostatic pressure. 
We speculate that, while the field quenches the fluctuations by enhancing the 
$XY$ anisotropy, the increasing pressure opposes this trend by canting the 
electronic in-plane orbitals, thereby, modifying the spin-density distribution.
Causes for the steady decrease of $T_{\mathrm{N}}$ with increasing pressure 
are, most likely, a small intrinsic in-plane rhombic distortion of the Cu-ion 
squares and a growing pressure-induced anisotropy. Both effects result in 
anisotropic angular deformations which oppose the AFM ordering and shift 
the 3D transition towards lower temperatures.

Even though the interlayer couplings induce a 3D AFM transition at nonzero 
temperatures, they do not significantly influence the trend of $T_{\mathrm{N}}$ 
under pressure. The main influence in tuning the transition temperature seems 
to be exerted by the planar deformation. It consists in the enhancement 
of the already documented weak in-plane frustration of \CUPZ, whereby the 
geometrical distortion of the square lattice at higher pressures modifies 
the relative weight of the NN and NNN terms in the Hamiltonian 
(Eq.~\ref{eq:Hamiltonian}), thus enhancing the quantum fluctuations which 
oppose the AFM order.

\begin{acknowledgments}
This work was financially supported in part by the Schweizerische Nationalfonds zur F\"{o}rderung der Wissenschaftlichen
Forschung (SNF) and the NCCR research pool MaNEP of SNF.
\end{acknowledgments}

\bibliography{Cupz_bibliography}

\end{document}